\documentclass[aps, twocolumn]{revtex4-1}
\usepackage{graphicx}
\usepackage{amssymb, amsmath}
\usepackage{bm}
\usepackage{hyperref}
\usepackage{color}

\newcommand{\Rb}[1]{$^ {#1}$Rb}
\newcommand{\gcorr}[2]{g^{\left( #1 \right)} \left(#2 \right)}

\definecolor{blue-comment}{rgb}{0.0, 0.5, 1.0}

\begin{document}

\title{Photon-correlation measurements of atomic-cloud temperature using an optical nanofiber}

\author{J. A. Grover}
\author{P. Solano}
\author{L. A. Orozco}
\author{S. L. Rolston}

\email{rolston@umd.edu}

\affiliation{Joint Quantum Institute, Department of Physics, University of Maryland and National Institute of Standards and Technology, College Park, MD 20742, USA}

\date{\today}

\begin{abstract}
We develop a temperature measurement of an atomic cloud based on the temporal correlations of fluorescence photons evanescently coupled into an optical nanofiber.
We measure the temporal width of the intensity-intensity correlation function due to atomic transit time and use it to determine the most probable atomic velocity, hence the temperature. 
This technique agrees well with standard time-of-flight temperature measurements. 
We confirm our results with trajectory simulations.
\end{abstract}

\maketitle


\section{Introduction}
\label{sec:intro}

Optical nanofibers (ONF) have emerged as a noninvasive probe for spectroscopy, sensing, and cold atom physics~\cite{Brambilla2010,Zhang2010,Garcia-Fernandez2011,Morrissey2013,Sague2007}.
In the case of cold atomic gases, the sub-$\mu$m diameter of the ONF allows for insertion into the atomic cloud with minimal disturbance to the trapping beams.
Moreover, the small mode area of the evanescent field around the ONF waist leads to strong coupling between the ONF guided mode and atoms near its surface~\cite{LeKien2006}.

Optical nanofibers have been used to study atomic spectra near surfaces~\cite{Sague2007,Russell2009,Nayak2012a} and magneto-optical trap (MOT) size, lifetime~\cite{Morrissey2009}, and temperature (by completely different methods than reported here)~\cite{Russell2012,Russell2013}.
Having a good temperature diagnostic is important, for example, in optimizing nanofiber trapping from a cold thermal cloud~\cite{LeKien2004,Vetsch2010,Goban2012,Beguin2014a,Lee2014}.
Standard techniques for measuring MOT temperature, such as time-of-flight (TOF) absorption imaging or dithering the magnetic field gradient~\cite{Russell2012}, are effective, but constrained experimental environments can prevent their use.

Hybrid quantum systems composed of superconductors and neutral atoms require cryogenic environments that are incompatible with many of the diagnostic tools used in laser cooling and trapping. 
To successfully interface cold atoms and superconducting qubits~\cite{Verdu2009,Hoffman2011,Xiang2013,Patton2013,Bernon2013,Jessen2014,Weiss2015} it is necessary to develop tests that do not perturb the cryogenic environment.
Superconductors are perilously sensitive to optical power and DC and AC magnetic fields.
The standard temperature measurements of the atomic cloud mentioned above are inaccessible in these setups due to a lack of optical access or sensitivity to changing magnetic fields.
Here we present, as part of our program to magnetically couple atoms trapped around an optical nanofiber to a superconducting resonator~\cite{Hoffman2011}, a new way to monitor the temperature of a cloud of cold atoms near a nanofiber, using the correlations of fluorescent photons emitted into the nanofiber guided mode. 

When the emitters are not stationary, the intensity-intensity correlation function depends on their motion as well as the geometry of the mode into which they emit~\cite{Carmichael1978,Kimble1978}.
The intensity-intensity correlation function, $g^{(2)}(\tau)$, measures correlations in the fluctuations of light intensity, e.g. the photon statistics~\cite{Loudon2000}, and can reveal both classical and quantum aspects of the light and its sources.
Among the classical effects is the transit time of atoms through an optical mode.
Systems such as atomic beams~\cite{Hennrich2005,Norris2009}, single atoms in a MOT~\cite{Gomer1998,Gomer1998a}, and a single trapped ion~\cite{Rotter2008} were used to measure these transit-time effects.
While bunched and antibunched photon statistics have been observed in the light emitted into the ONF guided mode~\cite{LeKien2008,Nayak2008,Nayak2009,Das2010}, the correlations related to atomic trajectories near the ONF have been studied only tangentially~\cite{Nayak2008}.
Intensity correlations decay with a characteristic time that depends on atomic transit. 
We measure this time for different atomic temperatures.
Its dependence on temperature allows for a simple model to infer the MOT temperature directly from the correlations.

This paper is organized as follows.
Section~\ref{sec:system} outlines the nanofiber mode structure, potential shifts, and the coupling efficiency of fluorescence photons into the nanofiber.
Section~\ref{sec:theory} provides a general overview of intensity-intensity correlations and briefly discusses the theoretical considerations for calculating and simulating them.
Finally Section~\ref{sec:expt} presents the experimental results and compares them to simulations.


\section{The system}
\label{sec:system}
The experiment relies on two main parts: a source of cold $^{87}$Rb atoms and an ONF.
A MOT provides a constant source of slowly moving atoms whose fluorescent light can couple evanescently into the guided mode of the optical nanofiber.
The nanofiber collects the light from the atoms and also modifies the local potential for the atoms, which move with typical velocities on the order of 10 $\mathrm{cm}\cdot\mathrm{s}^{-1}$.


\subsection{Nanofiber mode structure}
\label{subsec:mode}
Our single-mode nanofiber is a fiber pulled to a small enough diameter that all higher-order modes are cut off.
The mode (HE$_{11}$) of such an ONF has an intensity profile outside of the fiber given by~\cite{LeKien2004}
\begin{equation}
\lvert \mathbf{E}(r) \lvert^2 = \mathcal{E}^2\left[ K^2_0(qr) + u K^2_1(qr) + w K^2_2(qr)\right]\,,
\label{eq:intensity}
\end{equation}
where $\mathcal{E}^2$ is proportional to the intensity at the fiber surface; $K_i$ is the modified Bessel function of the second kind of order $i$; $u$ and $w$ are constants obtained from Maxwell's equations; $r$ is the distance from the center of the fiber; and $q = \sqrt{\beta^2-k^2}$ is the transverse component of the wavevector, where $\beta$ is the field propagation constant in the nanofiber, and $k = 2\pi/\lambda$ is the free-space wavevector.
The parameter $q$ describes the decay of the field in the radial direction.


\subsection{Atom-surface potential}
\label{subsec:potentials}
We approximate the nanofiber as an infinite dielectric plane when calculating the van der Waals potential~\cite{Alton2010,Stern2011,Frawley2012}, so that $U_{\mathrm{vdW}}(r)  = C_3 (r-r_0)^{-3}$ for $r>r_0$, where $r_0$ is the fiber radius.
The $C_3$ coefficient is equal to $4.94\times10^{-49}\,\mathrm{J}\cdot\mathrm{m}^{-3}$ and $7.05\times10^{-49}\,\mathrm{J}\cdot\mathrm{m}^{-3}$ for the $5 S_{1/2}$ and $5 P_{3/2}$ levels, respectively, of $^{87}$Rb near fused silica~\cite{LeKien2004,Lin2004,Sague2008a,Markle2014}.
The infinite-plane approximation is accurate to within 20\% for atom-fiber distances less than 200 nm~\cite{LeKien2004}, a distance comparable to the decay length of the evanescent field ($q^{-1} \approx 188$ nm, see Sec.~\ref{subsec:data}).

To include the effect of retardation, which causes the atom--surface interaction to scale as $(r-r_0)^{-4}$ for $r\gg r_0$, we use a phenomenological model for the potential that smoothly connects the non-retarded (van der Waals) and retarded (Casimir-Polder) regimes~\cite{LeKien2008b,Russell2009}:
\begin{equation}
U(r) = - \frac{C_4}{(r-r_0)^3 \left((r-r_0)+C_4/C_3 \right) }\,,
\label{eq:vdWCP}
\end{equation}
where $C_4$ is the Casimir-Polder coefficient.
This $C_4$ coefficient is equal to $4.47\times10^{-56}\,\mathrm{J}\cdot\mathrm{m}^{-4}$ and $12.2\times10^{-56}\,\mathrm{J}\cdot\mathrm{m}^{-4}$ for the $5 S_{1/2}$ and $5 P_{3/2}$ levels, respectively, of $^{87}$Rb near fused silica~\cite{Spruch1993,Sague2008a}


\subsection{Potential shifts}
\label{subsec:shifts}
The atom-surface potential shifts the atomic levels dependent on position.
The shifts produce a spatially-varying absorption (emission) rate~\cite{Foot2005}:
\begin{equation}
p_{\mathrm{abs}}\left( r \right) = \frac{\Gamma}{2} \frac{s}{1+s+4\left( \frac{d\omega \left( r \right) +\delta}{\Gamma} \right)^2}\,,
\label{eq:shift}
\end{equation}
where $r$ is the position of the atom, $s = I/I_{\mathrm{sat}}$ is the saturation parameter ($I_{\mathrm{sat}} = 3.58\, \mathrm{mW}\cdot\mathrm{cm}^{-2}$ for a uniform sublevel population distribution~\cite{Steck2001}), $\delta = \omega_{\mathrm{L}} - \omega_0$ is the detuning of the driving (i.e. MOT) beams from atomic resonance, and $d\omega \left( r \right) = \left( U_e(r) - U_g (r)\right ) / \hbar$ is the atom-surface shift assuming a two-level atom.

Note that we neglect effects due to interference of the MOT beams with each other or due to scattering of the MOT beams off the nanofiber in the near field.
This is justified because the fiber is long relative to these effects in one direction so that the atoms see a landscape that is, on average, uniform.


\subsection{Coupling efficiency}
\label{subsec:coupling}
The coupling efficiency of an atom to the ONF is the rate of spontaneous emission that couples into the one-dimensional mode of the fiber divided by the total spontaneous emission rate~\cite{LeKien2006,Masalov2013}, 
\begin{equation}
\beta \left ( r \right ) = \Gamma_{1\mathrm{D} }\left( r \right)/\Gamma_{ \mathrm{tot} } \left( r \right)\,.
\label{eq:coupling}
\end{equation}
Fermi's golden rule determines the form of $\beta \left ( r \right ) $, which follows the spatial variation of Eq.~\ref{eq:intensity}.

Photon detection in the experiment is a joint process of absorbing a photon from the MOT beams and emitting it into the nanofiber mode, which is mathematically described by the product of the photon emission rate in Eq.~\ref{eq:shift} and the coupling efficiency in Eq.~\ref{eq:coupling}.
It is the position-dependence of this joint probability that allows us to obtain information about the atomic motion.


\section{Correlations}
\label{sec:theory}

The intensity-intensity correlation function
\begin{equation}
g^{(2)}(\tau) = \frac{\left\langle I(t)\,I(t+\tau) \right\rangle}{\left\langle I(t) \right\rangle^2}\,,
\label{eq:corr}
\end{equation}
measures the conditional probability of detecting a photon at a time $\tau$ from recording another photon.
Here $\langle \cdot \rangle$ denotes an average over all $t$, and, in this discussion, $I(t)$ is the photocurrent or, equivalently, the photon counting rate at time $t$.
At its core, $g^{(2)}(\tau)$ characterizes the fluctuations in $I(t)$.
When measuring fluorescence from an atomic ensemble, the function contains contributions from different sources of fluctuations including single-atom field-field correlations, single-atom intensity-intensity correlations, different-atom field-field correlations, and different-atom intensity-intensity correlations.
Neglecting correlations between different atoms and assuming that they are motionless, we can write $g^{(2)}(\tau)$ as~\cite{Carmichael1978}
\begin{equation}
g^{(2)}(\tau) = 1+\left\lvert g_A^{(1)}(\tau) \right\rvert^2+\frac{1}{\bar{N}}g_A^{(2)}(\tau)\,,
\label{eq:g2}
\end{equation}
where $\bar{N}$ is the average atom number in a particular time window, and $g_A^{(2)}(\tau)$ and $g_A^{(1)}(\tau)$ are the single-atom intensity-intensity and field-field correlations, respectively.
For small atom number $\bar{N}$, we can observe the ``antibunching term'' $g_A^{(2)}(\tau)$.

Laser-cooled atoms are not stationary emitters.
The resonance fluorescence emitted into the fundamental mode exhibits correlations due to transit-time effects related to the geometry of that mode.
The atoms act as beacons signaling their position while passing near the nanofiber.
Accounting for the motion of atoms amounts to adding a temporal envelope $f(\tau)$ to Eq.~\ref{eq:g2}~\cite{Hennrich2005}, 
\begin{equation}
\label{eq:transitcorr}
g^{(2)}(\tau) = 1+ f(\tau) \lvert g_A^{(1)}(\tau) \rvert^2+ \frac{1}{ \bar{N} } f(\tau)g_A^{(2)}(\tau)\,.
\end{equation}
The function $f(\tau)$ generally depends on the environment and how the emitted light couples to the detection apparatus -- it is the shape of this temporal envelope that will allow us to extract information about the trajectories of atoms moving near an ONF.

We can relate the width of the correlation function $g^{(2)}(\tau)$ to the temperature of the atomic cloud by noting that the temperature determines the velocity distribution of the atoms and the speed of the atoms determines the timescale of the interaction with the nanofiber.
The ONF mode described by Eq.~\ref{eq:intensity} possesses a characteristic length scale of $1/q$.
Dividing this length by a characteristic speed of a Maxwell-Boltzmann distribution of atoms at a temperature $T$, which we take to be the most probable speed, $v_p = \sqrt{2 k_B T/m}$, yields a simple relationship between transit time and temperature:
\begin{equation}
\tau_0 = \frac{a}{q}\sqrt{\frac{m}{2 k_B T}}\,,
\label{eq:temp}
\end{equation}
where $a$ is an overall scale factor based on the geometry of the problem and on our choice of characteristic speed. We are not able to find an analytical form for $a$ from simple physical considerations, but used simulations to understand how geometric details affect $a$ (see Sec.~\ref{subsec:sim}).


\section{Experiment and results}
\label{sec:expt}


\subsection{Apparatus}
\label{subsec:app}

We load the MOT from the low-velocity tail of a background vapor of \Rb{87} atoms produced by a dispenser (see details in Ref.~\cite{Lee2013}).
We change the intensity and detuning of the cooling beams in order to controllably vary the temperature of the atomic cloud between $\sim160-840\,\mu \mathrm{K}$, as measured by time-of-flight expansion through fluorescence imaging.
Our ability to determine the atomic cloud temperatures is limited by the time-of-flight (TOF) imaging system in our setup combined with the low atom numbers for colder MOTs in steady state.
While we collected correlation data for ostensibly colder MOTs, we only present data for temperatures for which we could provide calibration to a known technique.
This is an indication that in certain circumstances the signal-to-noise ratio of the correlation measurement technique can be better than that of TOF.

The optical nanofiber (ONF) is produced via the flame brushing technique~\cite{Hoffman2014a,Birks1992}.
A hydrogen-oxygen flame acts as a local heat source to soften $125\,\mu \mathrm{m}$-diameter, single-mode fiber (Fibercore SM800) whose ends are pulled with computer-controlled linear motors.
This method reliably produces fibers of subwavelength diameters with transmission of the fundamental mode above 99\% and as high as 99.95\%, allowing them to sustain powers of hundreds of milliwatts in high vacuum~\cite{Hoffman2014a}.
Based on our fiber-pulling reproducibility, we know the transmission is greater than 95\%.
Relying on repeated, destructive measurements of the nanofiber diameter using a scanning electron microscope (SEM), we estimate the diameter of our ONF to be $530\pm 50$ nm, with a 1\% uniformity over a length of 7 mm.
This fiber diameter with the stated uncertainty accepts only one guided mode, described by Eq.~\ref{eq:intensity} above, at the experimentally relevant wavelength of 780 nm.

This same fiber has been in our apparatus for over two years with no noticeable degradation in quality. Rubidium atoms can coat the fiber surface and reduce transmission under operating pressures, but application of a thru-fiber heating beam with a power of more than a few $\mu$W is sufficient to desorb the atoms within a few ms.

We glue (EPO-TEK OG116-31) the fiber to a titanium u-shaped mount for stability, and attach the mount to a UHV-compatible manipulator system (VG Scienta Transax).
The manipulator consists of a motorized stepper motor along one axis and 2D manual translations stages along the other axes.
This manipulator works in conjunction with three pairs of magnetic shim coils to optimally overlap the nanofiber waist with the region of highest atomic density in the cloud.

\begin{figure}
\centering
\includegraphics[width=0.9 \columnwidth]{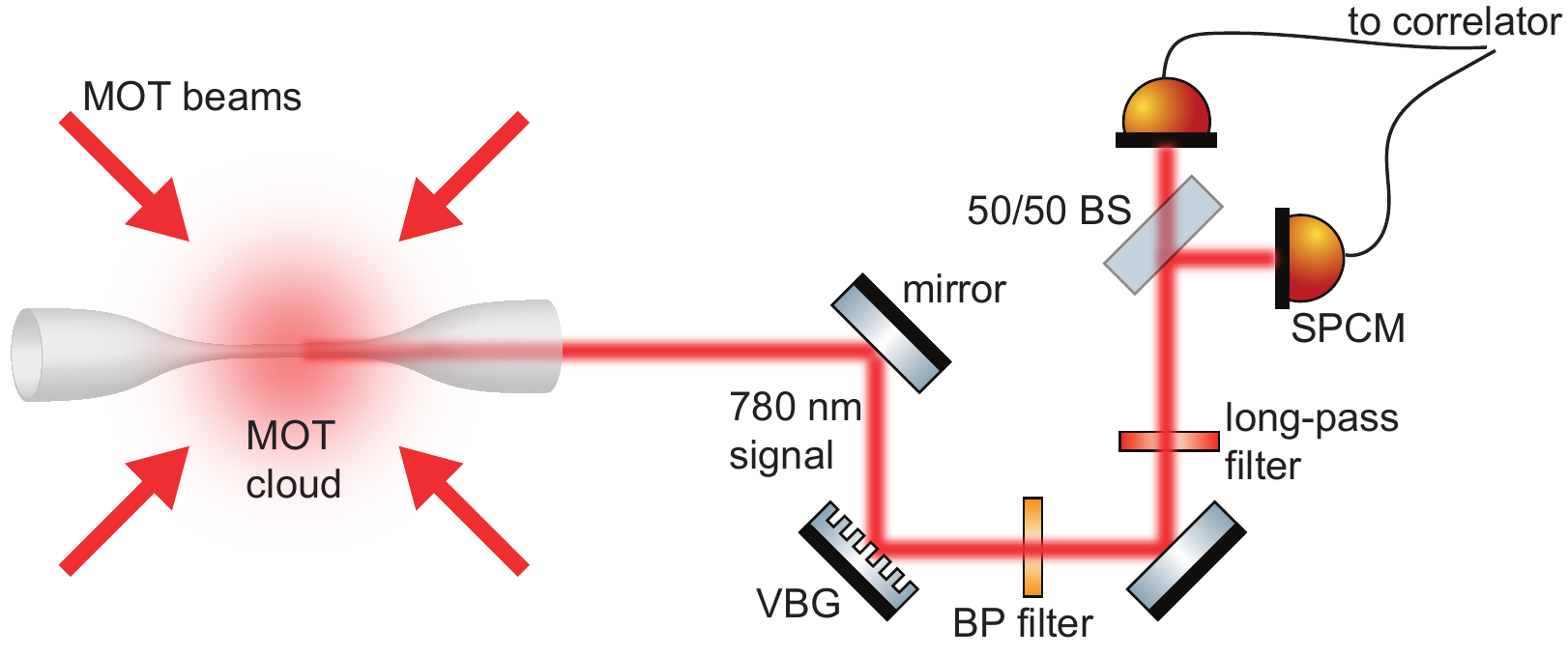}
\caption[Experimental schematic for correlation measurements]{\label{fig:schematic}(Color online) Experimental schematic. A MOT is spatially-overlapped with a nanofiber, and the MOT beams drive resonance fluorescence that couples into the guided mode. This signal is filtered by a volume Bragg grating (VBG), bandpass (BP) filter, and long-pass filter before being split by a 50:50 beamsplitter (BS) and sent to two SPCMs. TTL pulses from the SPCMs are time-tagged by an FPGA and correlated in software.}
\end{figure}

Light that couples into the guided mode is filtered at the output of the fiber by a volume Bragg grating (VBG, OptiGrate BP-785), a narrow-line interference filter (Semrock LL01-780-12.5), and a long-pass color filter (Thorlabs FGL645) before being sent to the two fiber-coupled single-photon counting modules (SPCMs) (see Fig.~\ref{fig:schematic}).
A field-programmable gate array (FPGA)~\cite{Peters2015} stores and time-tags photon output TTL pulses from the SPCMs, which are then post-processed and correlated.
An internal clock of 48 MHz sets the minimal time resolution to 20.83 ns.
The use of two SPCMs circumvents problems near zero time delay related to detector dead time, typically 50 ns. 

\subsection{Data and fitting}
\label{subsec:data}
For this experiment, the MOT beams are on continuously during data acquisition and drive fluorescent transitions in the atoms.
We collect $\sim 2.5\times 10^{7}$ photon counts for each experimental run, corresponding to about 45 min of averaging per data point.
Time-of-flight imaging measures the temperature of the atomic cloud before and after photon collection.

Our data is a list of times corresponding to photon detection events, which we use to find $g^{(2)}(\tau)$.
We do not do any further binning of the data, so that the timing resolution of 20.83 ns is set by the internal clock in the FPGA.
While this time resolution obscures details on atomic spontaneous emission timescales (tens of nanoseconds), it provides good resolution on the timescale of a few microseconds where the atomic trajectories produce signatures in the correlation function.
Measurements using an oscilloscope (Tektronix DPO 7054) with finer time resolution allowed us to observe antibunching for low atom number.

\begin{figure}
\centering
\includegraphics[width=0.9 \columnwidth]{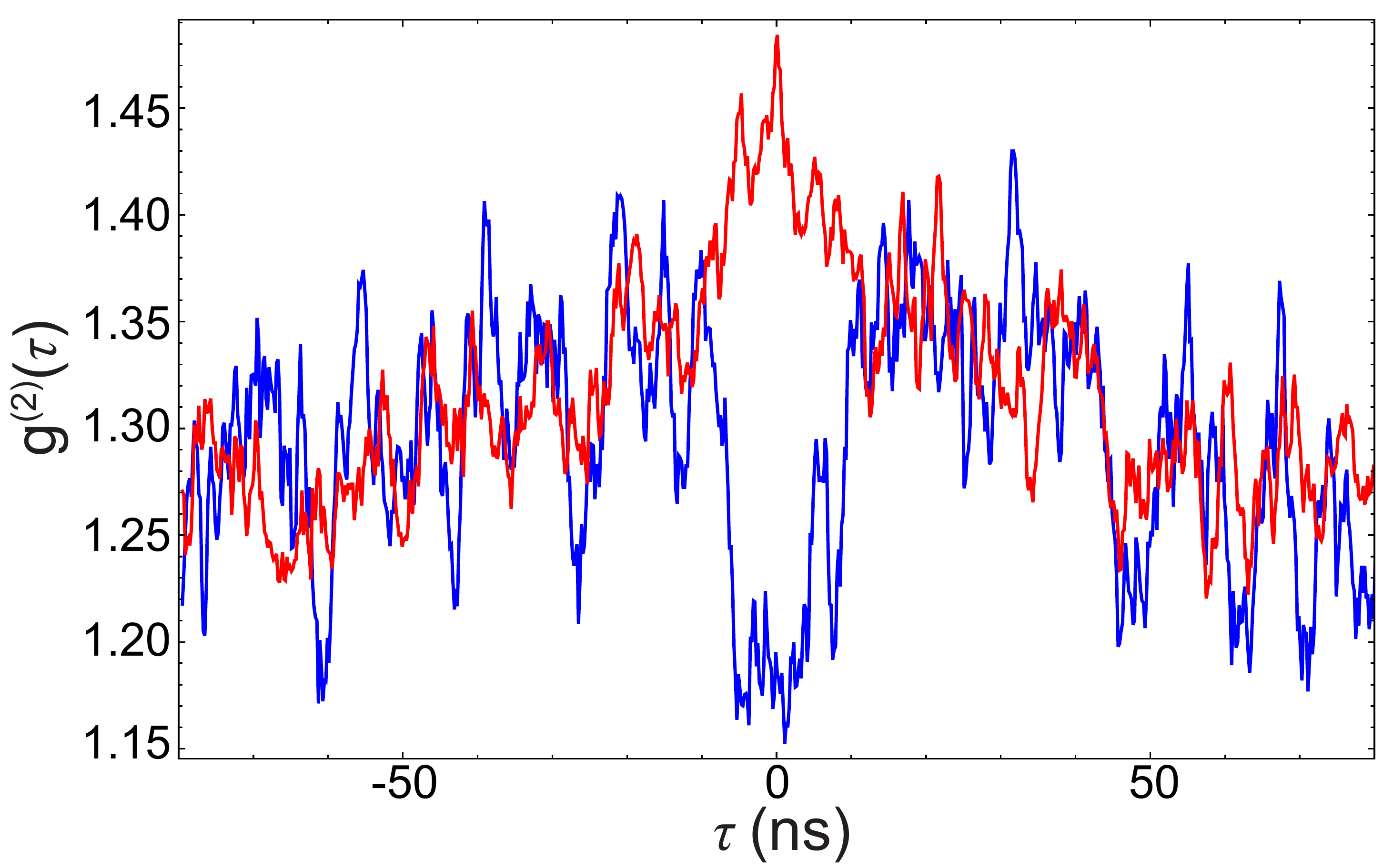}
\caption[Short timescale autocorrelation function of resonance fluorescence emitted into the nanofiber guided mode, illustrating a transition from bunching to antibunching]{\label{fig:antibunch}(Color online) Second-order correlation function $g^{(2)}(\tau)$ for light scattered into the fiber as a function of delay time $\tau$. 
The curves show data for low (blue) and high (red) Rb dispenser currents, illustrating antibunching and bunching, respectively.}
\end{figure}

Varying the Rb dispenser current allows us to change the number of atoms in the MOT, so that we can change the average number of atoms interacting with the nanofiber mode.
Fig.~\ref{fig:antibunch} shows the transition from antibunched (positive slope after $\tau=0$, estimated atom number is $\sim1.4$) to bunched (negative slope after $\tau=0$, estimated atom number is $\sim6$) correlations as we increase the number of atoms fluorescing into the mode of the ONF.
The timescale of the bunched or antibunched feature is set by the internal degrees of freedom of the atom and is much shorter than the temporal envelope due to atomic motion, which we discuss next.
Similar results were observed in Ref.~\cite{Nayak2009}.

\begin{figure}
\centering
\includegraphics[width=0.9 \columnwidth]{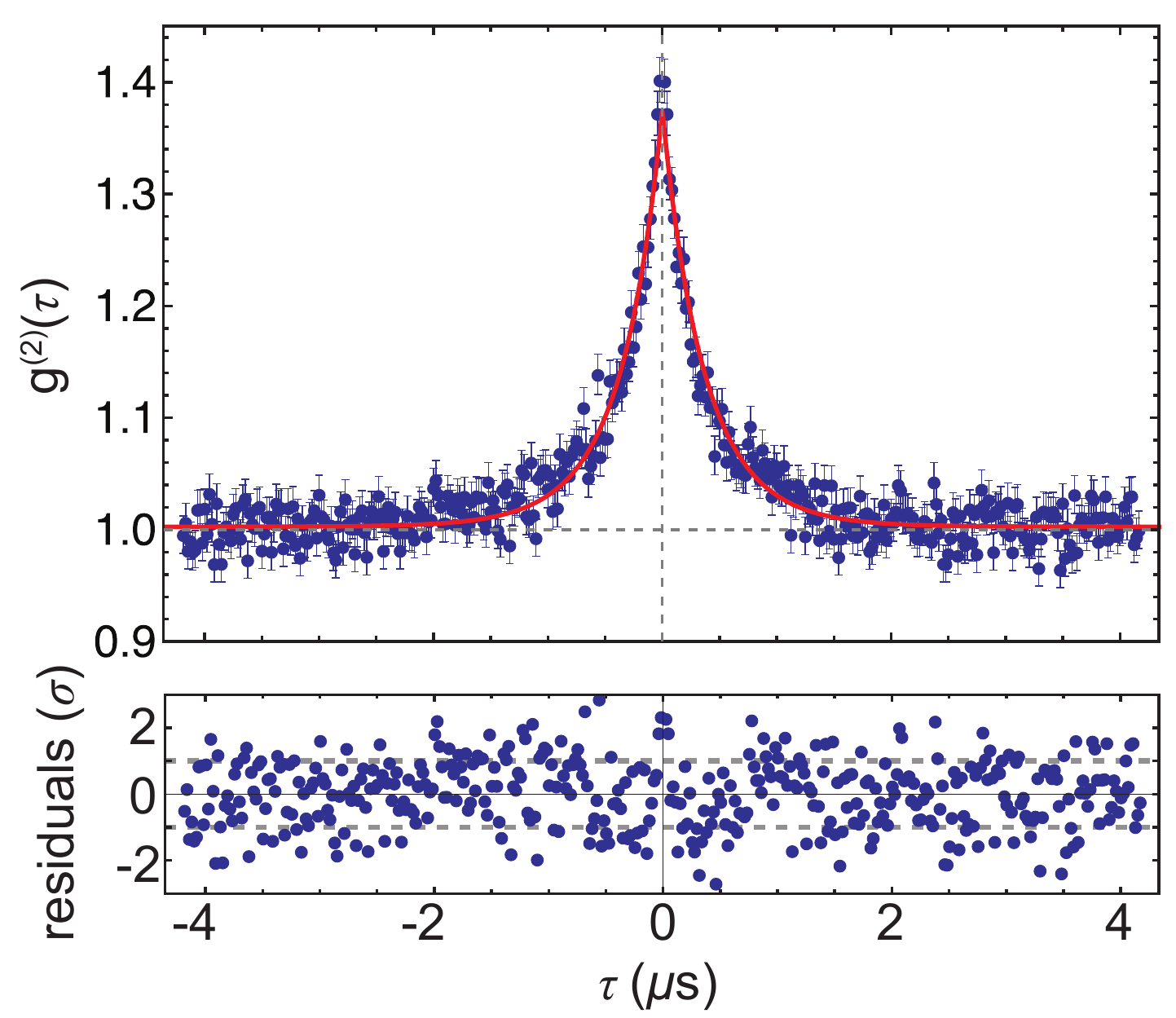}
\caption[Autocorrelation function that displays transit-time effects of atoms passing through the nanofiber guided mode, with fit and residuals]{\label{fig:fit}(Color online) Second-order correlation function $\gcorr{2}{\tau}$ as a function of delay time $\tau$ for an atom temperature of 460 $\mu$K. The data (blue dots) are fit (solid red line) to $1+f(\tau)$ using Eq.~\ref{eq:fit}, with the residuals displayed in the lower plot.}
\end{figure}

Figure~\ref{fig:fit} displays an example of $\gcorr{2}{\tau}$ extracted from data for an atom temperature of 460 $\mu$K.
Note the very different timescale than in Fig.~\ref{fig:antibunch} so that the atom-number-dependent peak or dip corresponds to only one data point in Fig.~\ref{fig:fit}.
Because this point at zero time delay is the only one that depends on atom number, we neglect it when fitting the data.
The signal has a characteristic width due to transit-time effects, which is the result of a position-dependent atom-fiber coupling efficiency combined with moving atoms.
An atom at a particular location will emit into the mode with probability proportional to its intensity at that position, and averaging over many atomic trajectories will sample the entire mode.
In this way, the autocorrelation function contains information about the mode in question (the shape of $\gcorr{2}{\tau}$) and about the motion of the atoms (the decay time of $\gcorr{2}{\tau}$).

We make a series of approximations to the model of the mode structure before comparing the observed transit-time broadening to theory.
The factors $u$ and $w$ in Eq.~\ref{eq:intensity} are small for a fiber radius of 265 nm and wavelength of 780.24 nm (calculated to be 0.166 and 0.00875, respectively), so we neglect them and keep only the first term, which is proportional to $K^2_0$.
As a further simplifying approximation we also take the asymptotic form of $K_{0}$~\cite{Olver2010},
\begin{equation}
K_{0} (z)\sim \sqrt{\frac{\pi}{2 z}}e^{-z} \,,
\label{eq:asymptote}
\end{equation}
which is valid in our case.
This yields an intensity around the nanofiber proportional to $\exp [-2qr]/2qr$.
Defining an effective index of refraction, $n_{\mathrm{eff}} = \beta/k$, we can rewrite the propagation constant so that the radial decay parameter becomes $q = k\sqrt{n^2_\mathrm{eff}-1}$, which is $0.66 k$ for our nanofiber.
We recast the spatial dependence of the intensity into the temporal envelope in Eq.~\ref{eq:transitcorr}~\cite{Hennrich2005,Norris2009}:
\begin{equation}
f(\tau) = A\, \frac{e^{-2\left ( \lvert \tau \rvert /\tau_0+ 0.66\, k\, r_0 \right )}}{\left ( \lvert \tau \rvert /\tau_0+ 0.66\,k\,r_0 \right )}\,,
\label{eq:fit}
\end{equation}
where $r_0 = 265$ nm is the fiber radius, $A$ is a fitting parameter for the overall amplitude, and the absolute value reflects the time-symmetric nature of the autocorrelation function for stationary processes.
The parameter $\tau_0$ represents a characteristic correlation time (see Eq.~\ref{eq:temp}).

The red curve in Fig.~\ref{fig:fit} shows the best fit to $g^{(2)}(\tau) = 1 + f(\tau)$ because $g^{(2)}_A (\tau )$ is flat in our experiment over timescales longer than the atomic lifetime.
For a measured atomic temperature of 460 $\mu$K, the fit achieves a reduced $\chi^2$ of 1.02 (and a range of approximately $1-1.5$ across all datasets).
We note that using Eq.~\ref{eq:fit} for the temporal envelope $f(\tau)$ results in statistically better fits than an exponential or Gaussian decay.

The overall height of the bunched peak depends on the absolute knowledge of the background at very long time (seconds), which we know can depend on mechanical vibrations of the fiber. 
The environment acoustically and thermally drives these vibrations so that we do not have an exact measure of unity.
This, combined with the signal-to-background ratio of the photon counting rates, can explain why the amplitude $A$ does not reach the expected value of 2 for chaotic light from independent emitters.


\subsection{$\tau_0$ vs. temperature}
\label{subsec:temp2}

\begin{figure} 
\centering
\includegraphics[width=0.9 \columnwidth]{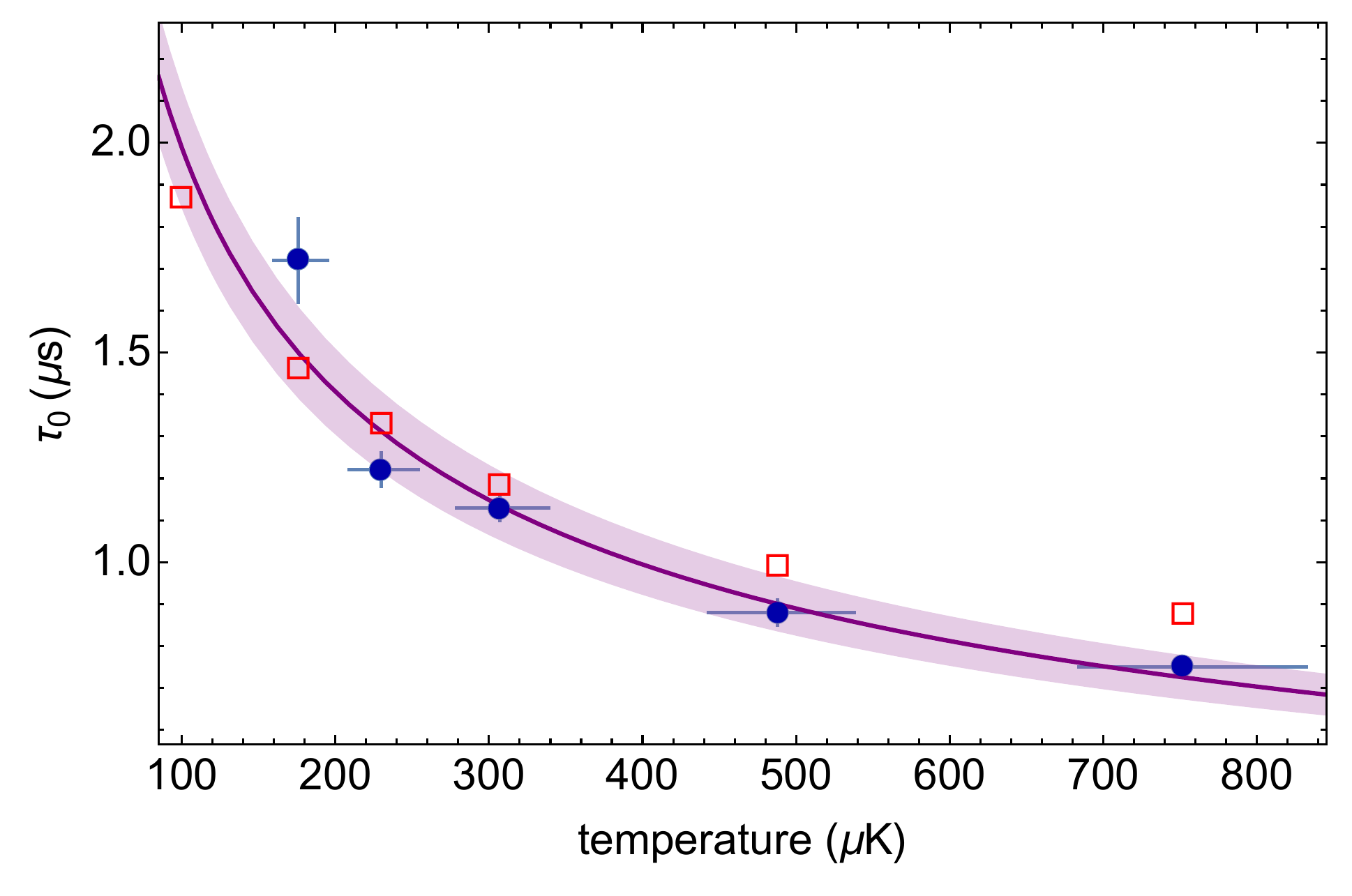}
\caption{\label{fig:temp}(Color online) Extracted $\tau_0$ vs. temperature $T$, measured via TOF. 
The blue circles are experimental data.
The vertical error bars indicate standard error in the fit of Eq.~\ref{eq:fit}, and the horizontal error bars arise from systematic uncertainty in the magnification of the imaging system. 
The purple line is a fit to Eq.~\ref{eq:temp}, and the shaded region is the 95\% ($2\sigma$) confidence bands considering both the vertical and horizontal error bars. 
The reduced $\chi^2$ is 1.67.
The red open squares are the results of the trajectory simulation, with a single scale parameter of 0.77.}
\end{figure}

We extract best-fit values for $\tau_0$ at different MOT atomic temperatures, with each temperature also measured by standard TOF imaging.
Fig.~\ref{fig:temp} shows a plot of the resulting best-fit values $\tau_0$, where the vertical error bars are the standard errors from the fit, and the horizontal error bars originate from uncertainty in the knowledge of the magnification of the imaging system.
The average atom number for all points in Fig.~\ref{fig:temp} falls in the same range as for the data presented in Fig.~\ref{fig:antibunch}, i.e. $1 < \bar{N} < 10$.
The purple line in Fig.~\ref{fig:temp} is a fit to Eq.~\ref{eq:temp} with the single fit parameter $a$, and the shaded area represents the 95\% ($2\sigma$) confidence bands considering both the vertical and horizontal error bars.
We observe good agreement between the model and the data; the fit has a reduced $\chi^2$ of 1.67, and the overall scale parameter is $a=1.46\pm0.04$.
The exact value of this scale factor is discussed further in Sec.~\ref{subsec:sim}.

\subsection{Simulations}
\label{subsec:sim}
To better understand the physical situation, we perform simulations of atomic trajectories subject to Newton's equations of motion~\cite{Sague2007}. 
These simulations include the atom-surface potential and its resultant shift discussed in Sections~\ref{subsec:potentials} and~\ref{subsec:shifts}.
The classical nature of the simulations is justified because the smallest angular momenta present in the system are still $\sim100$ times larger than $\hbar$.

The atoms are started at a radial distance $r=1500$ nm away from the fiber surface.
At this distance, the coupling is weak due to the rapid decay of the mode with length scale $1/q$.
The axial and radial symmetry of the problem allows us to restrict trajectories to the x-y plane with initial velocities pointing in one quadrant.
We sample the speeds from a 3D Maxwell-Boltzmann distribution before projecting onto this plane.
Trajectories evolve for either 50 $\mu$s or when the atom strikes the fiber surface, whichever happens first.

The coupling efficiency in Eq.~\ref{eq:coupling} is a fit to the complete solution for a two-level atom~\cite{Masalov2013}.
We also assume that the orientation of the atomic dipoles relative to the fiber surface is random, so that the coupling efficiency is an effective ensemble average. Independent measurements confirm that this assumption of random orientations is valid for our MOT.

Photon scattering events are infrequent on microsecond timescales, and when they do occur their effect on atomic velocity is negligible.
As a result, we assume ballistic trajectories.
The position-dependent coupling efficiency in Eq.~\ref{eq:coupling} and the position-dependent emission rate in Eq.~\ref{eq:shift} are calculated at each instant of time along a trajectory and multiplied together.
This yields a time-dependent detection probability for each trajectory.
Time-correlating the detection probability of a trajectory with itself produces a signal proportional to the intensity-intensity correlation for a single atom.
We discretize these time-dependent probabilities onto a mesh of 50-ns resolution so that calculating the correlation function becomes a simple array operation.
Experimentally-measured values for atom temperature are fed into the simulation, which is averaged over $5,000$ randomly sampled speeds and directions.
The resulting correlation function is fit to Eq.~\ref{eq:fit} in order to extract the decay time $\tau_0$.

\begin{figure}
\centering
\includegraphics[width=0.9 \columnwidth]{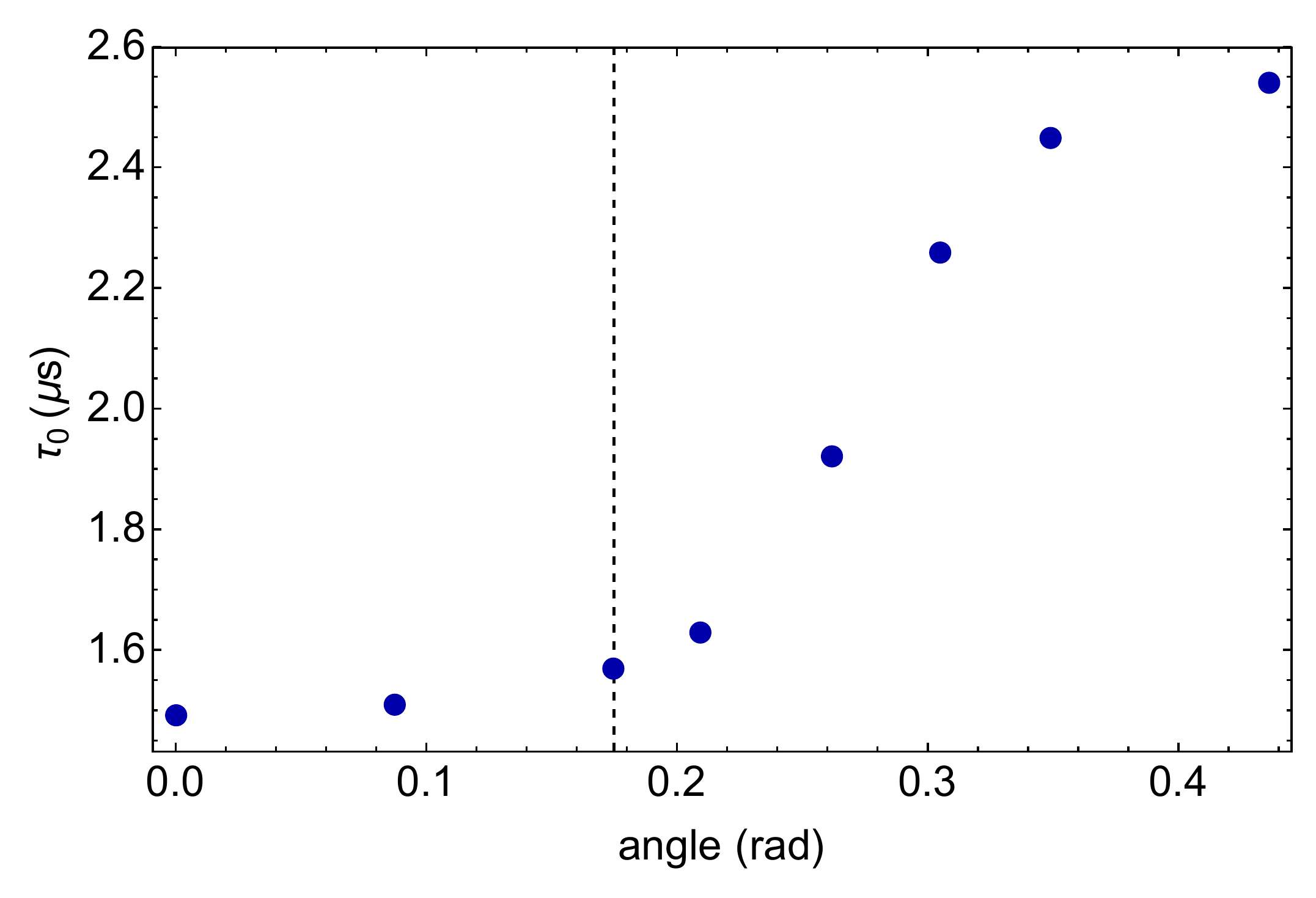}
\caption[Simulated correlation function width vs. sampling angle range for an atomic temperature of 90 $\mu$K]{\label{fig:simangle}(Color online) Simulated correlation time $\tau_0$ vs. sampling angle range $\Delta \theta$ for an atomic temperature of 90 $\mu$K. The dashed line indicates the critical angle $\arctan(53/300)$ in the simulation at which not all atoms hit the fiber.}
\end{figure}

We first utilize the simulations to address the scale factor in Eq.~\ref{eq:temp}.
Fig.~\ref{fig:simangle} displays the dependence of the transit time on the angular spread of the atomic trajectories for a distribution with temperature 90 $\mu$K.
For an atomic beam aimed directly at the fiber, we extract a transit time of 1.49 $\mu$s, which matches well the calculated time of 1.43 $\mu$s using Eq.~\ref{eq:temp} with $a=1$.
The transit time increases slowly as we increase the angular distribution of trajectories, until it hits a critical value of $\arctan(53/300)$, corresponding to the point after which not all paths intersect with the nanofiber.
Beyond this angle, atoms then interact with the fiber over distances longer than $1/q$, and the transit time consequently increases further.
Fig.~\ref{fig:simangle} illustrates that the simulation fully samples the interaction region with an angular spread of at least $\pi/6$ to get reasonable results.
Moreover, we note that the ratio of the transit time for the fully-sampled simulation to the effective one-dimensional simulation with no angular spread is 1.7.
These results suggest that our observed scale parameter of $a=1.46\,\pm\,0.04$ is partially due to angular spread in the trajectories.

We performed simulations for the same temperatures measured in the experiment and one additional atomic temperature of 100 $\mu$K.
The red open squares in Fig.~\ref{fig:temp} display transit times extracted from simulated data that were fit to the temporal envelope function, Eq.~\ref{eq:fit}.
The simulated data are multiplied by a single, overall scale parameter equal to 0.77 in order to minimize the least-squares distance to the experimental fit.
The simulations follow the expected trend with temperature and differ only by a scale factor of order unity.
The discrepancies might be explained by the various simplifications made in our model, which neglects, for instance, the stochastic nature of the photon absorption/emission process. 
We note, however, that the difference between the experimental and simulated data is comparable to other temperature measurement methods using optical nanofibers~\cite{Russell2013}.

 
\section{Conclusions}
\label{sec:concs}

We have presented a technique to measure the temperature of a laser-cooled atomic cloud that is applicable to experiments with restrictive environments, such as hybrid quantum systems using superconducting circuits.
The method uses the intensity-intensity correlation function to extract motion of atoms as they pass through the ONF mode and is easily extendable to other photonic devices with different optical mode geometries.
This technique allows mapping of mode structures, which could be useful when using the next family of higher-order modes to trap atoms near an optical nanofiber~\cite{Fu2007,Fu2008,Sague2008,Ravets2013,Kumar2015}.

\section{Acknowledgements}
This work was supported by the National Science Foundation through the PFC at JQI and the ARO Atomtronics MURI.
We gratefully acknowledge J. E. Hoffman for nanofiber fabrication, J. K. Peters for help with the FPGA for data acquisition, A. D. Cimmarusti for assistance with data processing software, H. J. Carmichael for discussions regarding simulations, and W. D. Phillips for a thoughtful reading of the manuscript.



%

\end{document}